%% file: morel.tex
\documentclass[mypaper,8pt,twoside]{CoAst}
\usepackage{epsf,graphicx,fancyhdr}
\input{CoAst_liege-proceedingslayo}

\def\rfr{\smallskip\par\noindent
        \hangindent=7truemm
        \hangafter=1}

\begin{document}
\sf

\chapterCoAst{Abundances of massive stars: some recent developments}
{Thierry Morel} 
\Authors{T.\,Morel} 
\Address{Institut d'Astrophysique et de G\'eophysique, Universit\'e de Li\`ege, All\'ee du 6 Ao\^ut, 4000 Li\`ege, Belgium}

\noindent

\begin{abstract}
Thanks to their usefulness in various fields of astrophysics (e.g. mixing processes in stars, chemical evolution of galaxies), the last few years have witnessed a large increase in the amount of abundance data for early-type stars. Two intriguing results emerging since the last reviews on this topic (Herrero 2003; Herrero \& Lennon 2004) will be discussed: (a) nearby OB stars exhibit metal abundances generally {\it lower} than the solar/meteoritic estimates; (b) evolutionary models of single objects including rotation are largely unsuccessful in explaining the CNO properties of stars in the Galaxy and in the Magellanic clouds.
\end{abstract}

\Session{\three} 

\section*{1 -- Introduction}
This review about the chemical properties of massive stars will focus on two issues that are relevant in the context of this meeting. First, one may wonder how the chemical abundances of nearby OB stars compare to the solar values. This piece of information is, for instance, needed to properly model B-type pulsators and to draw correct inferences about their internal structure. Second, the abundances of several elements are powerful probes of mixing phenomena and, as such, can be used to improve our theoretical understanding of these processes and ultimately better model the evolution of massive stars across the HR diagram. Fascinating physical phenomena such as mass-transfer processes can affect the abundances of stars in binaries, but we refrain from discussing these systems here (see Langer, these proceedings).

\section*{2 -- Getting the abundances}
A prerequisite to obtain reliable abundances is to adopt accurate atmospheric parameters (a favourable case is offered by detached eclipsing binaries; e.g. Pavlovski \& Southworth 2008). The effective temperature can be derived from photometric indices or, preferably, through ionisation balance of some metals (usually Si). The surface gravity is derived by fitting the collisionally-broadened wings of the Balmer lines, while the microturbulent velocity is inferred by requiring the abundances of a given ion to be independent of the line strength. Model atmospheres (either LTE or NLTE) with adequate line-blanketing are required. Departures from LTE are significant in hot stars and a full NLTE treatment for the line formation is also needed. Plane-parallel (e.g. TLUSTY, DETAIL/SURFACE) or so-called unified codes (e.g. CMFGEN, FASTWIND) can be used depending on the strength of the stellar wind. Of course, more sophisticated analysis techniques (NLTE model atmosphere, spherical extension)  are much more demanding in terms of computer resources and should be used with discernment. A hybrid approach involving hydrostatic, LTE model atmospheres coupled with an NLTE line-formation treatment is often employed for early B-type stars on the main sequence (e.g. Nieva \& Przybilla 2008). At this stage, a set of model atoms as complete as possible should have also been developed. Contrary to the situation for cool stars where one can calibrate the $\log gf$ values using a solar spectrum obtained with the same instrumental set up, for hot stars one has to rely instead on theoretical calculations. The technique used for the abundance determination (classical curve-of-growth vs spectral synthesis) evidently depends on the seriousness of blending issues and hence on the $v \sin i$ of the star under study. As one might expect, most studies are strongly biased towards narrow-lined stars (see below).

\section*{3 -- Chemical composition of nearby OB stars: solar?}
To estimate the baseline chemical properties of massive stars in the solar neighbourhood, abundance data from the literature have been gathered using the following criteria: (a) known binaries are avoided; (b) supergiants are omitted to minimise evolutionary effects (they will be discussed in the following in the context of mixing); (c) only stars within $\sim$1 kpc are discussed because of the existence of a Galactic metal gradient (e.g. Daflon \& Cunha 2004). Only NLTE studies based on high-resolution spectra have been selected, while carbon abundances solely based on transitions not well modelled by most line-formation codes (e.g. C II $\lambda$4267) have been excluded. Note that some bright, easily-observable stars can be represented more than once. As seen in Fig.\ref{fig1}, most of these $\sim$150--200 stars are slowly-rotating, early B-type dwarfs/subgiants (data from Cunha \& Lambert 1994, Daflon \& Cunha 2004 and references therein, Gies \& Lambert 1992, Gummersbach et al. 1998, Kilian 1992, 1994, Lyubimkov et al. 2005, Morel et al. 2008, Nieva \& Przybilla 2008 and Thompson et al. 2008).

\figureCoAst{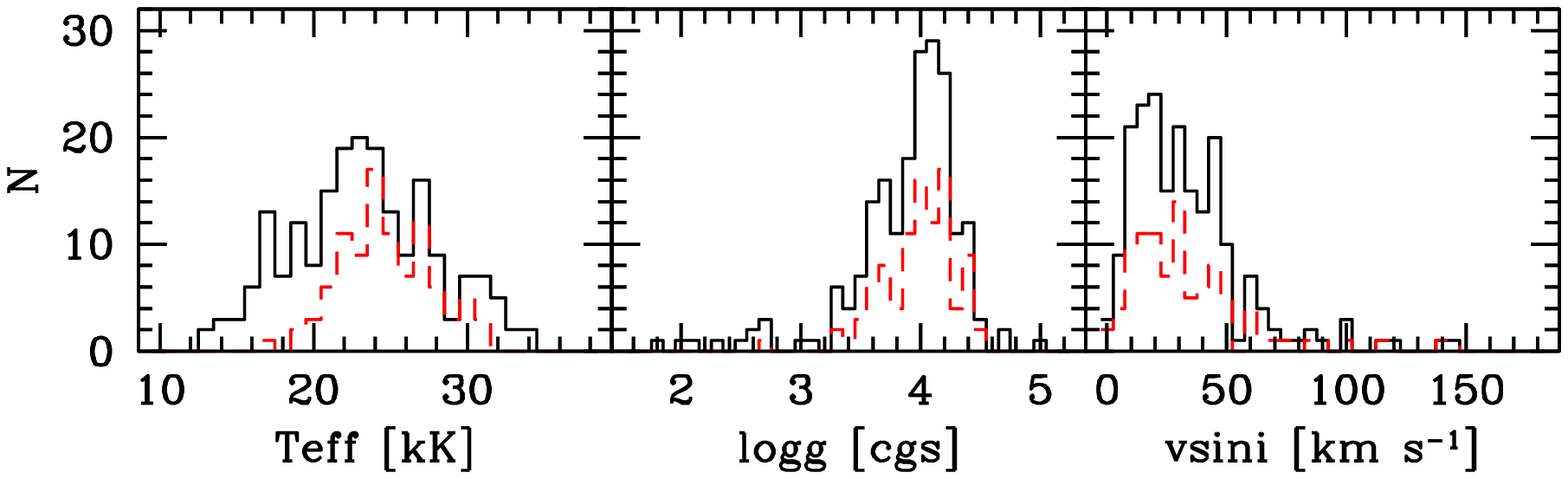}{Distribution of the physical parameters for nearby OB stars with abundance data (either C, N, O, Mg, Al, Si, S or Fe; {\it solid line}) and CNO data ({\it dashed line}).}{fig1}{h!}{clip,angle=0,width=97mm}

\figureCoAst{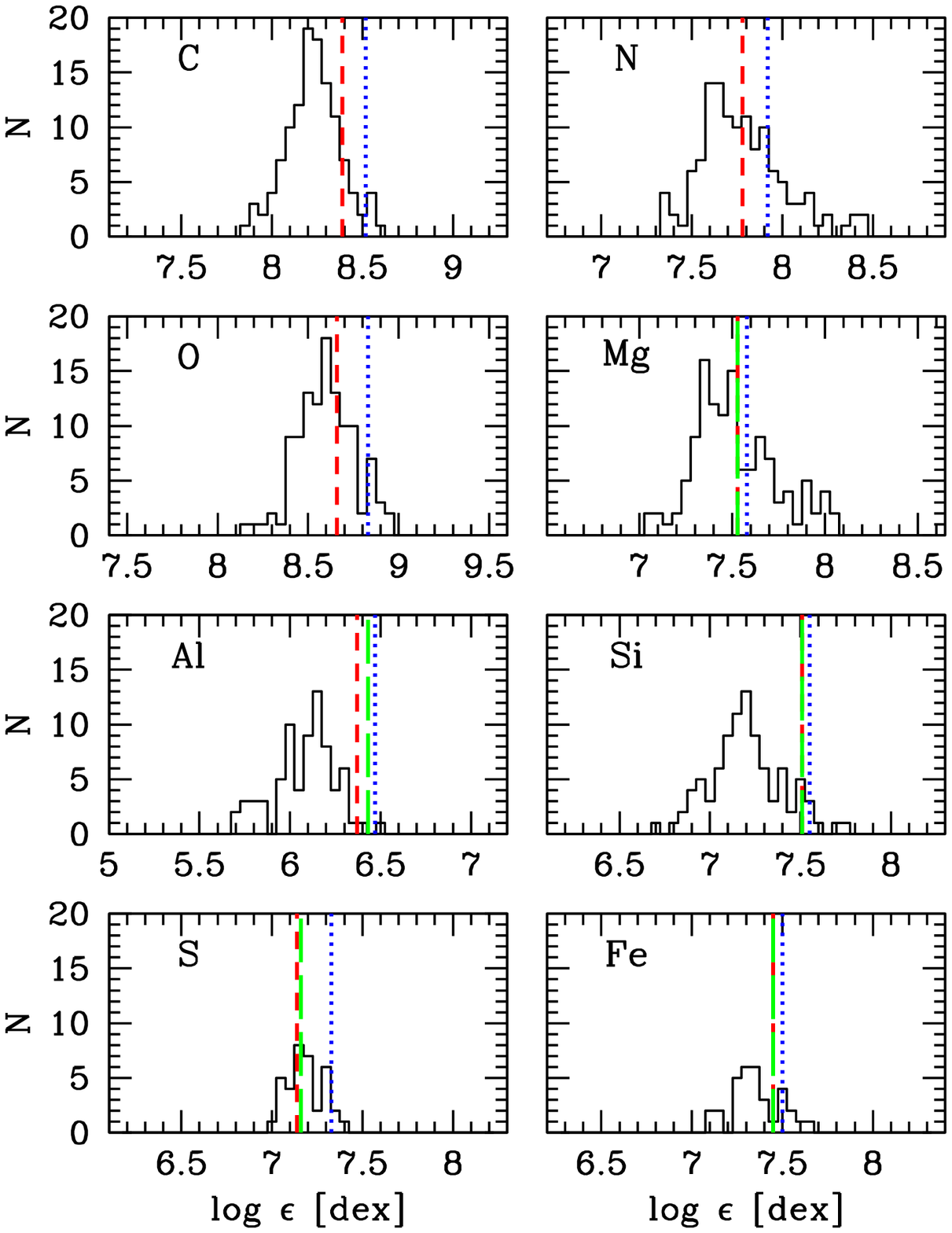}{Distribution of the metal abundances of nearby OB stars. The logarithmic abundances are given on a scale in which $\log \epsilon$(H)=12. {\it Dotted line}: solar abundances of Grevesse \& Sauval (1998), {\it short-dashed and long-dashed lines}: solar and meteoritic abundances of Asplund et al. (2005), respectively.}{fig2}{h!}{clip,angle=0,width=82mm}

Before comparing the abundances of young OB stars with the solar values, two remarks are necessary. First, chemical evolution models of the Galaxy predict a small enrichment of the local ISM in metals over the past 4.6 Gyr (typically $\sim$0.1 dex; Chiappini et al. 2003) and the solar photosphere is slightly depleted in metals because of gravitational settling (typically $\sim$0.05 dex; Turcotte et al. 1998). Nearby, unevolved OB stars are hence naturally expected to appear metal rich compared to the Sun, but this should only be at moderate levels. Second, the solar abundances are indistinguishable from the mean values obtained for nearby, early G stars of the Galactic thin disk analysed in exactly the same way (Allende-Prieto 2006). The Sun therefore does not appear peculiar in terms of bulk metallicity or elemental abundances (despite hosting giant planets) and its chemical composition should be representative of the one prevailing in the local ISM at the time of its formation (see also, e.g. Gustafsson 2008).
 
As can be seen in Fig.\ref{fig2}, the mean abundances of several metals (especially C, Al and Si) in  nearby OB stars are found to be significantly lower than the solar values derived either using classical model atmospheres (Grevesse \& Sauval 1998) or 3-D hydrodynamical simulations (Asplund et al. 2005). Although the data for OB stars are highly inhomogeneous, this is a result found by most studies (if not all in case of certain elements). The results of Kilian (1992, 1994) do not appear clearly discrepant despite the fact that they are based on model atmospheres that are not fully line blanketed. Low abundances are also consistently found for Mg, but the distribution depicted in Fig.\ref{fig2} is strongly biased by the much higher values (up to 0.5 dex above solar) found by Lyubimkov et al. (2005).\footnote{A number of figures showing the results of the individual studies are available from: {\tt http://www.astro.ulg.ac.be/$\sim$morel/liege\_colloquium.pdf}} Nitrogen deserves special comment, as the larger spread observed is likely the consequence of deep mixing already on the main sequence (as will be discussed below). The concomitant C and O depletions are of smaller amplitude and are not expected to be readily detectable at these levels of N enrichment. Two NLTE studies are available in the literature for both S and Fe. In the former case, solar values were found (Daflon \& Cunha 2004; Morel et al. 2008). For Fe, however, both subsolar (Morel et al. 2008) and solar values have been reported (Thompson et al. 2008). The fact that the mean Al, Si and Fe abundances are lower than the accurate meteoritic values supports the idea that the abundances of nearby OB stars are generally underestimated (by up to a factor 2 in the case of Si), perhaps as a result of missing physics or unaccounted systematic errors. Such a discrepancy is hard to explain on theoretical grounds (although some models have been proposed; see, e.g. Witt 2001). The neon abundance of nearby B stars is consistent with the estimate of Grevesse \& Sauval (1998) based on coronal observations of the Sun and does not support the high theoretical value required to restore the past agreement between the standard solar models and the helioseismic constraints (Morel \& Butler 2008, and references therein). Although the argon abundance of B stars is much higher than the solar estimate (Lanz et al. 2008), the values for these two noble gases are in excellent agreement with those inferred in the ionised gas of the Orion nebula (Esteban et al. 2004).

\section*{4 -- Deep mixing in OB stars}
\subsection*{Mixing on the main sequence and close to it}
Figure \ref{fig3} shows the CNO logarithmic abundance ratios for the stars discussed above. The [N/C] ratio, which is a robust indicator of CNO-processed material dredged up to the surface, shows evidence for two subsamples of stars with either roughly solar values ([N/C]$\sim$--0.5 dex) or significantly higher ratios ([N/C]$\sim$--0.1 dex). The latter population is unexpected considering that these stars are core-hydrogen burning objects with low $v\sin i$ values (Fig.\ref{fig1}). The [N/O] data do not clearly show evidence for two distinct populations, but rather suggest a continuum of values with a tail extending to high ratios (see also Herrero \& Lennon 2004). The N-rich stars tend to be slightly more evolved than the N-normal stars ($\Delta \log g$$\sim$0.25 dex; Fig.\ref{fig4}), suggesting that this nitrogen excess could arise from an evolutionary effect (however, the unphysically large gravities reaching up to $\log g$=4.6 dex found for many N-normal stars cast some doubt on this interpretation; see also Fig.\ref{fig6} where these stars fall well below the ZAMS). On the other hand, a higher incidence of an N excess in magnetic B stars compared to stars without a field detection is emerging (Morel et al. 2008), but more work is needed to firmly establish such a (probably statistical) link between an N enhancement and magnetic fields. Abundance data for large samples of both magnetic and  non-magnetic stars are in particular needed to clearly reveal a dichotomy between the two groups. In any case, it is interesting to note in the context of models incorporating magnetic fields that some N-rich stars possess a large-scale, dipole field most likely of fossil origin (e.g. $\zeta$ Cas, $\beta$ Cep).

\figureCoAst{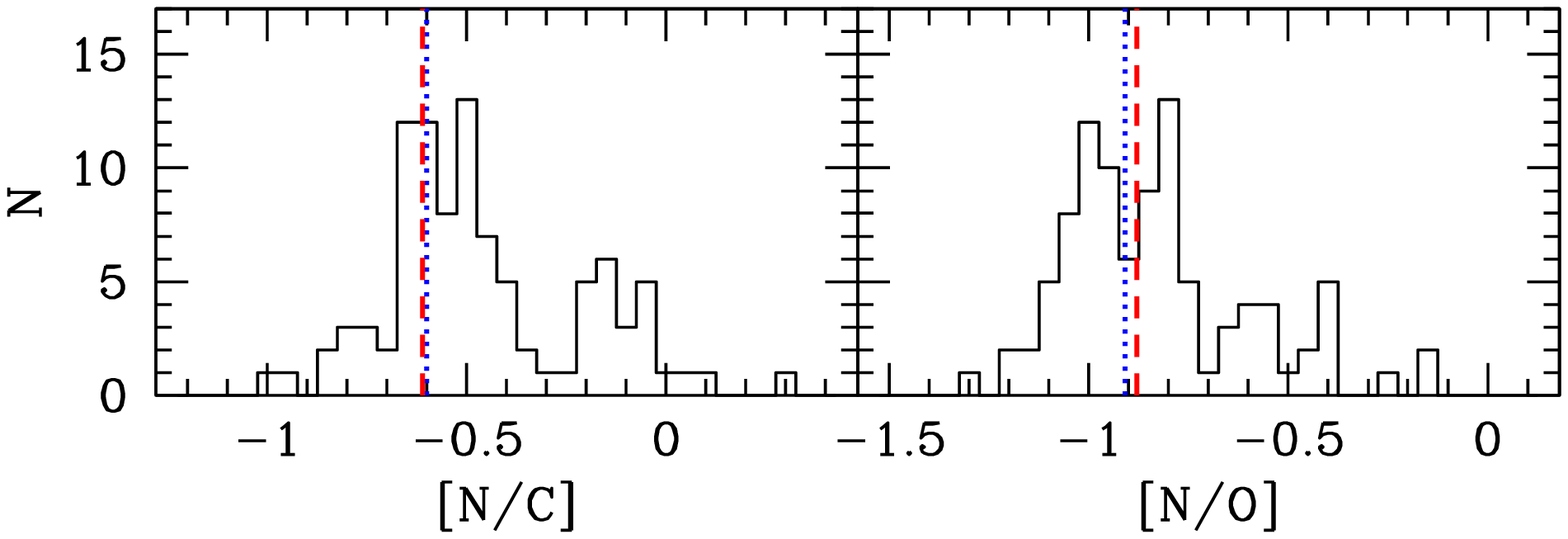}{Distribution of the [N/C] and [N/O] ratios for nearby OB stars. The {\it dotted} and {\it short-dashed} lines indicate the solar ratios of Grevesse \& Sauval (1998) and Asplund et al. (2005), respectively.}{fig3}{h!}{clip,angle=0,width=86mm}

\figureCoAst{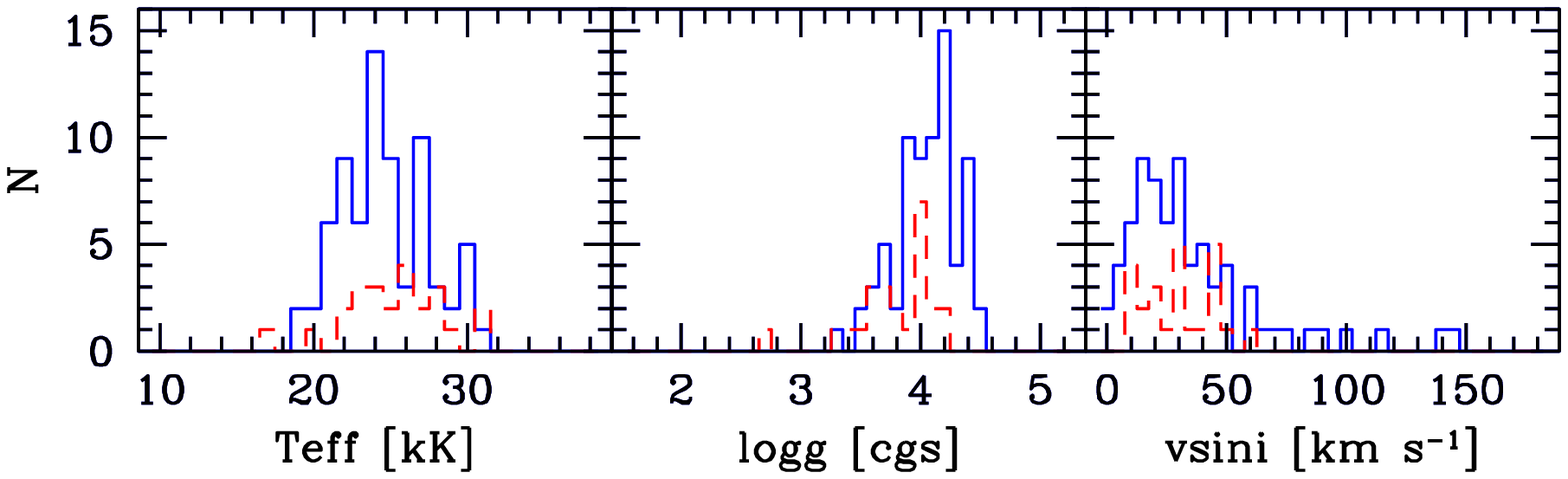}{Distribution of the physical parameters for the N-normal ([N/C]$<$--0.3 dex; {\it solid line}) and N-rich stars ([N/C]$>$--0.3 dex; {\it dashed line}).}{fig4}{h!}{clip,angle=0,width=97mm}

To examine whether this population of slowly-rotating dwarfs with an N excess is predicted by theory, we compare in Fig.\ref{fig5} the abundance data with the predictions of evolutionary models including rotational effects (Heger \& Langer 2000). The N-rich stars have not yet evolved beyond the TAMS (Fig.\ref{fig4}), and rotational velocities reaching $\sim$200 km s$^{-1}$ on the ZAMS are needed to reproduce their CNO properties. This can be contrasted with their low projected, present values ($<$$v\sin i$$>$$\sim$30 km s$^{-1}$; Fig.\ref{fig4}). Some $\beta$ Cephei stars in this sample have been shown to be {\it intrinsically} very slowly rotating based on seismic studies (Morel et al. 2006). Assuming that the N-rich stars were rapid rotators on the ZAMS, but then dramatically spun down to the observed levels, is not supported by models (e.g. Meynet \& Maeder 2003) or observations (e.g. Huang \& Gies 2006) that only suggest a modest loss of angular momentum in this mass range during core-hydrogen burning. Magnetic braking is also unlikely to strongly spin down B stars with a dipole field at the few hundred Gauss level (ud-Doula et al. 2008). 

\subsection*{Mixing in evolved objects}
A broader and more complete picture of the incidence of deep mixing across the HR diagram can be gained by considering all stars irrespective of their evolutionary status or Galactocentric distance (we complement the previous data with results from Crowther et al. 2006, Kilian et al. 1994, Kilian-Montenbruck et al. 1994, Mathys et al. 2002, Przybilla et al. 2006, Schiller \& Przybilla 2008, Searle et al. 2008, Smiljanic et al. 2006, Venn 1995 [with updated N abundances from Venn \& Przybilla 2003], Villamariz et al. 2002, Villamariz \& Herrero 2005 and Vrancken et al. 2000). The variation of the [N/C] abundance ratio across the $\logg$-$\log T_{\rm eff}$ plane clearly reveals evolutionary effects in the sense that all supergiants are N-rich by up to two orders of magnitude compared to main-sequence stars of similar mass (Fig.\ref{fig6}). The extreme enhancements observed in several cool supergiants may result from the first dredge up phase. As discussed above, two populations of normal and mildly-enriched N stars with similar masses coexist between the ZAMS and the TAMS. There is a tentative indication that the N-rich stars cluster at lower $v\sin i$ values (Fig.\ref{fig7}; see also Fig.\ref{fig4}), but it is unfortunately not yet possible to thoroughly test rotational mixing theories on Galactic objects as only very few CNO data for fast rotators are available. It is hoped that the data for fast-rotating stars in three Galactic open clusters obtained in the course of the VLT/FLAMES survey of massive stars will soon fill this caveat. Data for the SMC/LMC are, however, already available and have interestingly revealed two populations that cannot be accounted for by rotational mixing operating in single objects (Hunter et al. 2008; Brott, these proceedings): slowly-rotating dwarfs with an N excess (as in the Galaxy) and fast-rotating stars with normal nitrogen.

\figureCoAst{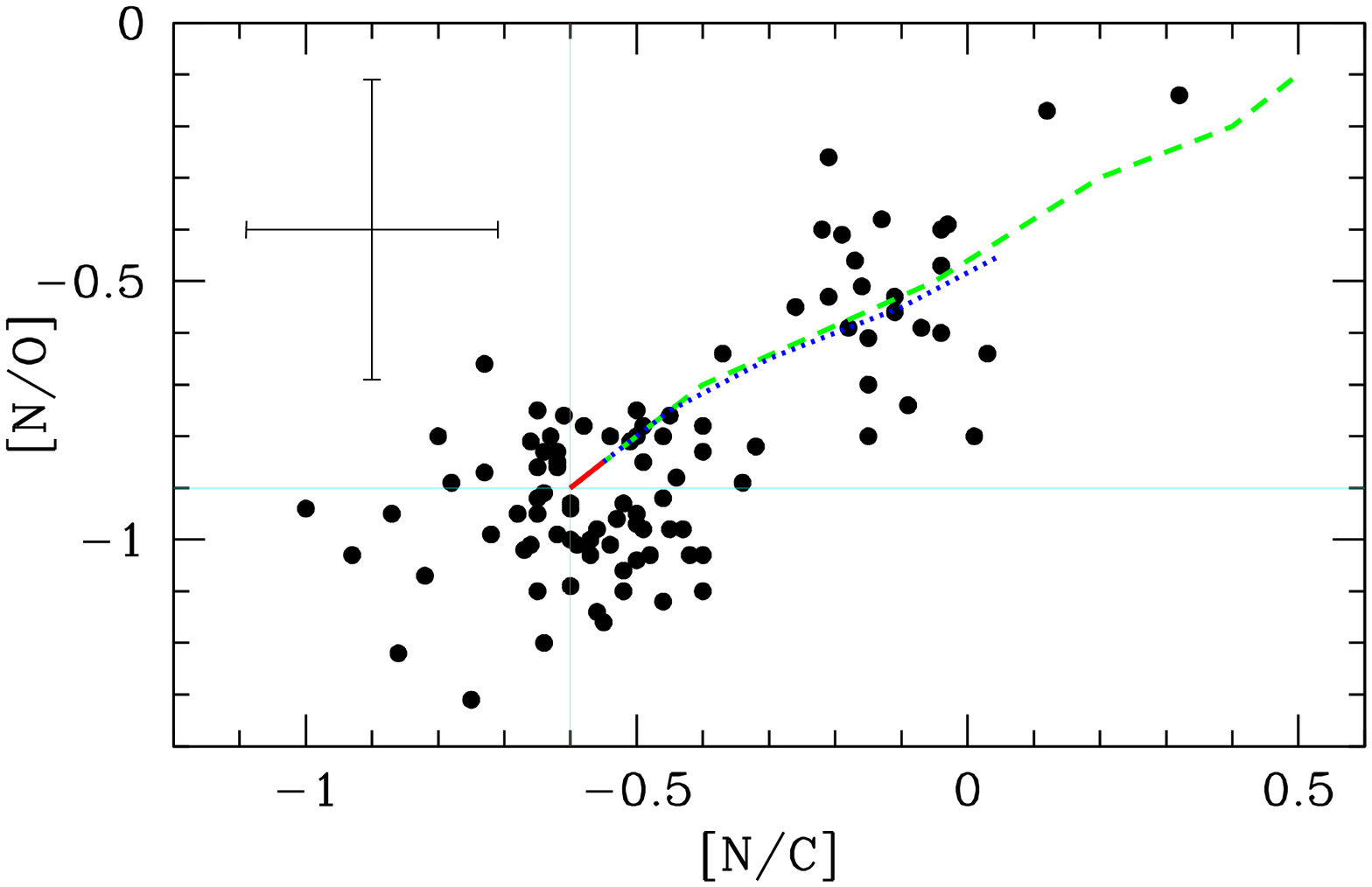}{Comparison between the observed [N/C] and [N/O] ratios of nearby OB stars and the predictions of evolutionary models including rotation (Heger \& Langer 2000). The results are shown for a 12 M$_{\odot}$ star (representative of this sample) and three values of the rotational velocity on the ZAMS: 99 ({\em solid line}), 206 ({\em dotted line}) and 328 km s$^{-1}$ ({\em dashed line}). The loci define an age sequence with time increasing rightwards from ZAMS to TAMS. The solar ratios are indicated by horizontal and vertical solid lines.}{fig5}{h!}{clip,angle=0,width=74mm}

\section*{6 -- Summary and directions for future research}
Most abundance studies are strongly biased towards early B-type stars. It would be desirable in the future to have more data for O stars. Unexpected and exciting results are also likely for late B-type stars because of the existence of diffusion effects in this $T_{\rm eff}$ range, but very few systematic studies have been undertaken up to now (e.g. Hempel \& Holweger 2003). 

The metal abundances of nearby OB stars are found to be in most cases below the most recent (and likely realistic) estimates for the Sun or, more importantly, the meteoritic values. The fact that the abundances (after correction for dust depletion) of C, N, O and S in the ionised gas of the Orion nebula (Esteban et al. 2004) are in good agreement with the new solar values lends credence to the idea that the abundances of massive stars are generally underestimated and seems to rule out scenarios such as a recent infall of metal-poor material at the Sun location as an explanation for the low abundances of these objects. We should point out, however, that the values closer to solar very recently reported by Przybilla et al. (2008) for a small sample of early B-type stars suggest that improvements in the data analysis (e.g. better model atoms and/or temperature scale) could help solving this discrepancy. 

The surface abundances of $\beta$ Cephei stars are indistinguishable from the values found for (presumably) non-pulsating B0--B3 main-sequence stars, but some objects are also N-rich and hence display the signature of deep mixing (Morel et al. 2006). Taking the results for B stars at face value would imply a higher {\it relative} abundance of the iron-peak elements compared to the solar mixture, and therefore that the pulsation modes are more easily excited (Montalb{\'a}n et al., these proceedings). Until the discrepancy with the solar/meteoritic values discussed above is better understood, however, a sound assumption would be to use the solar mixture of Asplund et al. (2005) in oscillation codes. As discussed during this meeting, one of the priorities for the future is to incorporate the detailed abundances of the SMC/LMC in such theoretical codes (and not simply a scaled solar pattern) in order to account for the existence of B-type pulsators in such low-metallicity environments and to adequately model their pulsation properties. Such a work is underway (Salmon et al., in preparation).

Several elements are powerful probes of mixing processes (He, B, CNO). There have been claims of an increase of the He abundance along the main sequence (Lyubimkov et al. 2004; Huang \& Gies 2006), but it is still unclear whether this effect is real or is merely an artefact of the analysis (e.g. choice of microturbulence). On the other hand, boron provides precious (and complementary) information about shallow mixing close to the surface (e.g. Mendel et al. 2006). Clear evolutionary effects are observed, with the supergiants all exhibiting high [N/C] ratios. Recent observations have revealed the existence of two populations challenging the very importance of rotational mixing as mechanism dredging up material from the convective core to the photosphere in massive stars: (a) slowly-rotating, N-rich dwarfs (mixing efficiency underestimated, stronger loss of angular momentum than expected and/or magnetic fields?); (b) fast-rotating stars with normal nitrogen in the LMC (binaries?). Establishing the nature of these two populations is needed for further progress. This implies that more effort should be devoted to the daunting and arduous task of deriving accurate abundances for fast rotators.

\figureCoAst{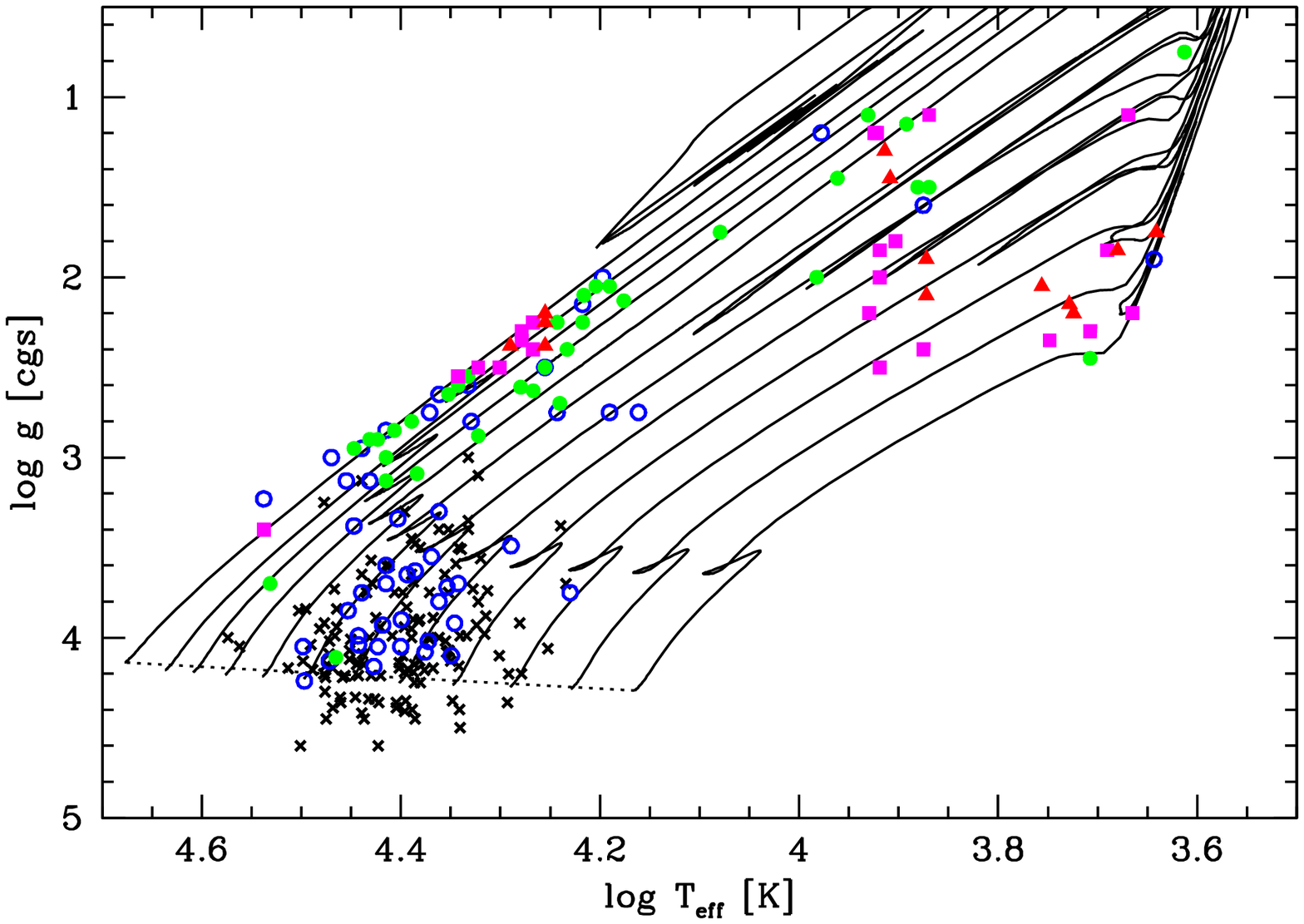}{Variation of the [N/C] abundance ratio across the $\logg$-$\log T_{\rm eff}$ plane (the evolutionary tracks without rotation and for $M$=4.0, 5.0, 6.3, 7.9, 10.0, 12.6, 15.8, 20.0, 25.1, 31.6, 39.8 and 63.1 M$_{\odot}$ are taken from Claret 2004). {\it Crosses}: [N/C]$<$--0.2, {\it open circles}: --0.2$<$[N/C]$<$+0.2, {\it filled circles}: +0.2$<$[N/C]$<$+0.6, {\it filled squares}: +0.6$<$[N/C]$<$+1.0, {\it filled triangles}: [N/C]$>$+1.0 dex.}{fig6}{h!}{clip,angle=0,width=97mm}

\figureCoAst{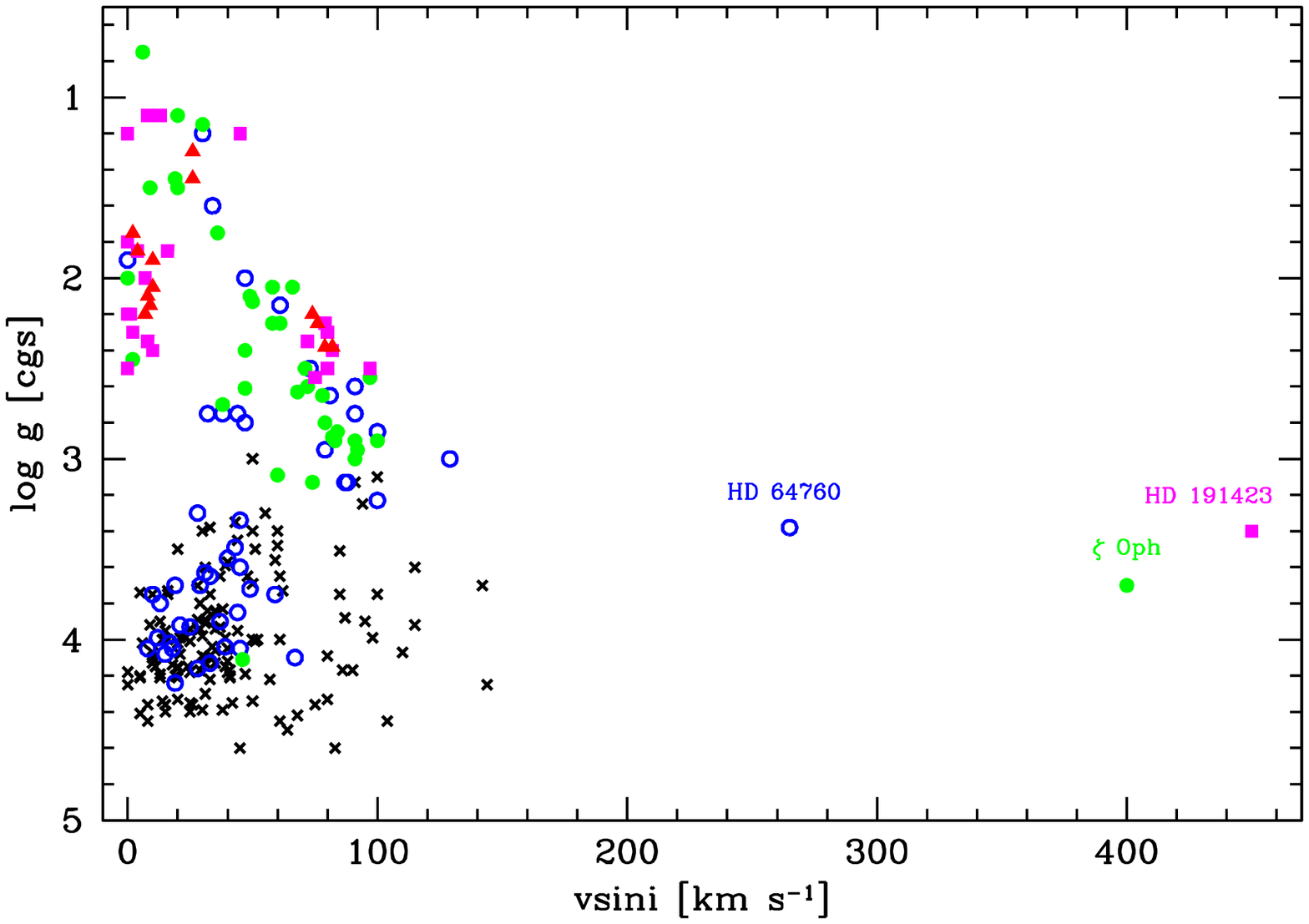}{Variation of the [N/C] abundance ratio across the $\logg$-$v\sin i$ plane (symbols as in Fig.\ref{fig6}). The three fast rotators have been analysed by Searle et al. (2008), Villamariz et al. (2002) and Villamariz \& Herrero (2005). Note that stars with widely different masses are shown in this figure.}{fig7}{h!}{clip,angle=0,width=97mm}

\References{
\rfr Allende-Prieto, C. 2006, in The Metal-Rich Universe, in press (astro-ph/0612200)\\
\rfr Asplund, M., Grevesse, N., \& Sauval, A. J. 2005, in Cosmic abundances as records of stellar evolution and nucleosynthesis, ed. T. G. Barnes III, F. N. Bash, ASP Conf. Ser., 336, 25\\
\rfr Chiappini, C., Romano, D., \& Matteucci, F. 2003, MNRAS, 339, 63\\
\rfr Claret, A. 2004, A\&A, 424, 919\\
\rfr Crowther, P. A., Lennon, D. J., \& Walborn, N. R. 2006, A\&A, 446, 279\\
\rfr Cunha, K., \& Lambert, D. L. 1994, A\&A, 426, 170\\
\rfr Daflon, S., \& Cunha, K. 2004, ApJ, 617, 1115\\
\rfr Esteban, C., Peimbert, M., Garc\'{\i}a-Rojas, J., et al. 2004, \mnras, 355, 229\\
\rfr Gies, D. R., \& Lambert, D. L. 1992, ApJ, 387, 673\\
\rfr Grevesse, N., \& Sauval, A. J. 1998, Space Sci. Rev., 85, 161\\ 
\rfr Gummersbach, C. A., Kaufer, A., Sch\"{a}fer, D. R., et al. 1998, A\&A, 338, 881\\  
\rfr Gustafsson, B. 2008, Phys. Scr., T130, 014036\\
\rfr Heger, A., \& Langer, N. 2000, ApJ, 544, 1016\\
\rfr Hempel, M., \& Holweger, H. 2003, A\&A, 408, 1065\\ 
\rfr Herrero, A. 2003, in CNO in the Universe, ed. C. Charbonnel, et al., ASP Conf. Ser., 304, 10\\
\rfr Herrero, A., \& Lennon, D. J. 2004, in Stellar rotation, ed. A. Maeder, P. Eenens, 209\\
\rfr Huang, W., \& Gies, D. R. 2006, \apj, 648, 591\\
\rfr Hunter, I., Brott, I., Lennon, D. J., et al. 2008, ApJ, 676, L29\\
\rfr Kilian, J. 1992, A\&A, 262, 171\\
\rfr Kilian, J. 1994, A\&A, 282, 867\\
\rfr Kilian, J., Montenbruck, O., \& Nissen, P. E. 1994, A\&A, 284, 437\\
\rfr Kilian-Montenbruck, J., Gehren, T., \& Nissen, P. E. 1994, A\&A, 291, 757\\
\rfr Lanz, T., Cunha, K., Holtzman, J., \& Hubeny, I. 2008, ApJ, 678, 1342\\ 
\rfr Lyubimkov, L. S., Rostopchin, S. I., \& Lambert, D. L. 2004, MNRAS, 351, 745\\
\rfr Lyubimkov, L. S., Rostopchin, S. I., Rachkovskaya, T. M., et al. 2005, MNRAS, 358, 193\\
\rfr Mathys, G., Andrievsky, S. M., Barbuy, B., et al. 2002, A\&A, 387, 890\\
\rfr Mendel, J. T., Venn, K. A., Proffitt, C. R., et al. 2006, ApJ, 640, 1039\\
\rfr Meynet, G., \& Maeder, A. 2003, \aap, 404, 975\\
\rfr Morel, T., Butler, K., Aerts, C., et al. 2006, A\&A, 457, 651\\  
\rfr Morel, T., Hubrig, S., \& Briquet, M. 2008, A\&A, 481, 453\\
\rfr Morel, T., \& Butler, K. 2008, A\&A, 487, 307\\
\rfr Nieva, M. F., \& Przybilla, N. 2008, A\&A, 481, 199\\
\rfr Pavlovski, K., \& Southworth, J. 2008, MNRAS, submitted\\
\rfr Przybilla, N., Butler, K., \& Becker, S. R. 2006, A\&A, 445, 1099\\
\rfr Przybilla, N., Nieva, M. F., \& Butler, K. 2008, ApJ, 688, L103\\
\rfr Schiller, F., \& Przybilla, N. 2008, A\&A, 479, 849\\
\rfr Searle, S. C., Prinja, R. K., Massa, D., et al. 2008, A\&A, 481, 777\\  
\rfr Smiljanic, R., Barbuy, B., de Medeiros, J. R., et al. 2006, A\&A, 449, 655\\
\rfr Thompson, H. M. A., Keenan, F. P., Dufton, P. L., et al. 2008, MNRAS, 383, 729\\
\rfr Turcotte, S., Richer, J., Michaud, G., et al. 1998, ApJ, 504, 539\\
\rfr ud-Doula, A., Owocki, S. P., \& Townsend, R. H. D. 2008, MNRAS, in press (arXiv:0810.4247)\\
\rfr Venn, K. A. 1995, ApJ, 449, 839\\
\rfr Venn, K. A., \& Przybilla, N. 2003, in CNO in the Universe, ASP Conf. Ser., 304, 20 \\
\rfr Villamariz, M. R., Herrero, A., Becker, S. R., et al. 2002, A\&A, 388, 940\\
\rfr Villamariz, M. R., \& Herrero, A. 2005, A\&A, 442, 263\\
\rfr Vrancken, M., Lennon, D. J., Dufton, P. L., et al. 2000, A\&A, 358, 639\\
\rfr Witt, A. N. 2001, Phil. Trans. R. Soc. Lond. A, 359, 1949
}

\end{document}

%% file: CoAst_liege-proceedingslayo.tex
\renewcommand{\chaptermark}[1]{\markboth{#1}{}}

\newcommand{\apj}{ApJ}
\newcommand{\aap}{A\&A}  
\newcommand{\mnras}{MNRAS}

\newcommand{\logg}{\ensuremath{\log g}}                     

\newcommand{\kopf}{\small\itshape Comm. in Asteroseismology \\ Contribution to the Proceedings of the 38$^{th}$\,LIAC\,/\,HELAS-ESTA\,/\,BAG, 2008
}

\newcommand{\Authors}[1]{\begin{center}\normalsize\bf\sf #1 \end{center}}

\renewcommand{\author}[1]{\begin{center}\normalsize\bf\sf #1 \end{center}}
\newcommand{\Address}[1]{\begin{center}\small\sf #1 \end{center}}

\DeclareGraphicsExtensions{.eps,.jpg}

\newcommand{\Session}[1]{{\vspace{3mm}\small \noindent  \hspace*{3mm} Session: } #1 \normalsize}

	\newcommand{\three}{\small Atmospheres, mass loss and stellar winds}

\renewenvironment{abstract}{\section*{Abstract}\normalsize\sf}{}
\newcommand{\References}[1]{\begin{flushleft}{\large References\\}\vspace*{2mm}\small #1 \end{flushleft}}

\newcommand{\chapterCoAst}[2]{\chapter[\sf\normalsize #1\\ \footnotesize \hspace*{5mm}by #2 \sf\normalsize][]{#1\\}\rhead[\fancyplain{}{\sf\footnotesize \center{#1}}]{\fancyplain{}{\sffamily\thepage}}\lhead[\fancyplain{\kopf}{\sffamily\thepage}]{\fancyplain{\kopf}{\sf\footnotesize \center{#2}}}}

\newcommand{\figureCoAst}[5]{\begin{figure}[#4]
\centering
\includegraphics*[#5]{#1}
\caption{#2}
\label{#3}
\end{figure}}

\renewcommand{\chaptername}{}
